\documentclass[fleqn]{llncs}

\usepackage{latexsym}
\usepackage[dvips]{graphicx}
\usepackage{subfigure}
\usepackage{amssymb}
\usepackage{amsmath}
\usepackage{algorithm}
\usepackage{algorithmic}
\usepackage{longtable}
\usepackage{caption}

\usepackage{rotating}
\usepackage{color}
\definecolor{blank}{rgb}{0.7,0.7,0.7}
\definecolor{gray}{rgb}{0.5,0.5,0.5}

\newcommand{\biblio}[1]{./Biblio.bib}

\makeatletter
\newenvironment{prog}{\vspace{1.0ex}\par
\obeylines\@vobeyspaces\tt}{\vspace{1.0ex}\noindent } \makeatother
\newcommand{\startprog}{\begin{prog}}
\newcommand{\stopprog}{\end{prog}\noindent}

\def\defemb#1#2{\expandafter\def\csname #1\endcsname
                              {\relax\ifmmode #2\else\hbox{$#2$}\fi}}

\DeclareSymbolFont{symbolsC}{U}{pxsyc}{m}{n}
\SetSymbolFont{symbolsC}{bold}{U}{pxsyc}{bx}{n}
\def\re@DeclareMathSymbol#1#2#3#4{%
    \let#1=\undefined
    \DeclareMathSymbol{#1}{#2}{#3}{#4}}
\re@DeclareMathSymbol{\dashleftrightarrow}{\mathrel}{symbolsC}{101}
\defemb{cA}{{\cal A}}
\defemb{cB}{{\cal B}}
\defemb{cC}{{\cal C}}
\defemb{cD}{{\cal D}}
\defemb{cE}{{\cal E}}
\defemb{cF}{{\cal F}}
\defemb{cG}{{\cal G}}
\defemb{cH}{{\cal H}}
\defemb{cI}{{\cal I}}
\defemb{cJ}{{\cal J}}
\defemb{cL}{{\cal L}}
\defemb{cM}{{\cal M}}
\defemb{cN}{{\cal N}}
\defemb{cO}{{\cal O}}
\defemb{cP}{{\cal P}}
\defemb{cR}{{\cal R}}
\defemb{cS}{{\cal S}}
\defemb{cT}{{\cal T}}
\defemb{cU}{{\cal U}}
\defemb{cV}{{\cal V}}
\defemb{cX}{{\cal X}}
\defemb{cZ}{{\cal Z}}

\long\def\comment#1{}

\begin{document}
%==============================================================
%PORTADA INFORMES DE INVESTIGACIÃN DSIC-II
%==============================================================

\begin{titlepage}

\begin{large}

    \centerline{DEPARTAMENTO DE SISTEMAS INFORM\'ATICOS Y COMPUTACI\'ON}
    \centerline{UNIVERSITAT POLIT\`ECNICA DE VAL\`ENCIA}
    \centerline{}
    \centerline{P.O. Box: 22012 \quad\quad\quad E-46071 Valencia (SPAIN)}

\vspace{1cm}
\begin{center}
   \resizebox{4cm}{!}{\includegraphics{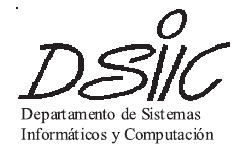}}
\end{center}
\vspace{1cm}

\centerline{\Huge{\bf\sffamily Informe T\'ecnico~/~Technical Report}}
\vspace{\baselineskip}\hrule\hrule

\vspace{3cm}

\begin{center}
\begin{tabular}{|lp{12cm}|}\hline
&\\
{\bf Ref. No.: } & 2022/01\hspace{3cm} {\bf Pages:} 13\\
{\bf Title: } &  A Benchmark Suite for Template Detection and Content Extraction\\
{\bf Author(s): } &  J. Alarte and J. Silva\\
{\bf Date:} & July, 2022\\
{\bf Keywords: } & Template Detection, Content Extraction, Benchmark Suite, Information Retrieval\\
& \\\hline
\end{tabular}
\end{center}

\vspace{2cm}
\begin{center}
\begin{tabular}{c@{\hspace{3cm}}c}
V$^\text{o}$ B$^\text{o}$\\
Leader of research Group & Author(s)\\
\end{tabular}
\end{center}

\end{large}

\end{titlepage}
\cleardoublepage
\pagestyle{myheadings}
\setcounter{page}{1}

\title{A Benchmark Suite\\ for Template Detection and Content Extraction\thanks{
This work has been partially supported by the EU (FEDER) and the
Spanish \emph{Ministerio de Econom\'{\i}a y Competitividad
(Secretar\'{\i}a de Estado de Investigaci\'on, Desarrollo e Innovaci\'on)}
under grant TIN2013-44742-C4-1-R and by the
\emph{Generalitat Valenciana} under grant PROMETEO/2011/052.
}}

%   jalarte@dsic.upv.es, 2 dinsa@dsic.upv.es, 3 jsilva@dsic.upv.es Departamento de Sistemas Informa?ticos y Computacio?n Universitat Polite`cnica de Vale`ncia, Valencia, Spain
%4 stamarit@babel.ls.fi.upm.es
%Babel Research Group
%Universidad Politcnica de Madrid, Madrid, Spain

  \author{%
  Juli\'an Alarte \and Josep Silva}

\institute{
Universitat Polit\`ecnica de Val\`encia, Camino de Vera S/N,
E-46022 Valencia, Spain. \\
\email{\{jalarte,jsilva\}@dsic.upv.es}
}

\maketitle

\begin{abstract}
Template detection and content extraction are two of the main areas of information retrieval applied to the Web. 
They perform different analyses over the structure and content of webpages to extract some part of the document. However, their objective is different. While template detection identifies the template of a webpage (usually comparing with other webpages of the same website), content extraction identifies the main content of the webpage discarding the other part. Therefore, they are somehow complementary, because the main content is not part of the template.   
It has been measured that templates represent between 40\% and 50\% of data on the Web. Therefore, identifying templates is essential for indexing tasks because templates usually contain irrelevant information such as advertisements, menus and banners. Processing and storing this information is likely to lead to a waste of resources (storage space, bandwidth, etc.). Similarly, identifying the main content is essential for many information retrieval tasks. 
Furthermore, there are other interesting block detection techniques such as menu detection, which tries to isolate the main menu of the webpage.
In this paper, we present a benchmark suite to test different approaches for template detection and content extraction. 
The suite is public, and it contains real heterogeneous webpages that have been labelled so that different techniques can be suitable (and automatically) compared.
\end{abstract}
%\vspace{-1em}
%\hspace{1cm}\textbf{Keywords:} Benchmark suite, Content Extraction, Template Detection, Information Retrieval.

%%%%%%%%%%%%%%%%%%%%%%%%%%%%%%%%%%%%%%%%%%%%%%%%%%%%%%%%%%%%%%%%%%%%%%%%%%%%%%%%

\section{Introduction}

Template extraction is an important tool for website developers, and also for website analyzers such as crawlers. 
Content extraction is essential for many information processing tasks applied to webpages. 
In the last decade, there have been important advances that produced several techniques for both disciplines. 
Hybrid methods that exploit the strong points of several techniques have been defined too. 
In order to test, compare and tune these techniques, researchers need:

\begin{itemize}
\item collections of benchmarks that are heterogeneous (to ensure generality of the techniques) and
\item a gold standard (to ensure the same evaluation criteria).
\end{itemize}  

A benchmark suite is essential to measure the performance of these techniques, and to compare them with previous approaches. Benchmark suites are used in the testing phase and in the evaluation phase. 
The testing phase allows developers to optimize the techniques by adjusting parameters. 
Once the technique has been tuned, the evaluation phase allows us to know its performance with objective measures. 
It is obvious that the set of benchmarks used in the testing phase cannot be used in the evaluation phase, thus, they need disjoint sets of webpages.

In this paper we present a benchmark suite together with a gold standard that can be used for template detection, content extraction and menu detection. All benchmarks have been labelled so that every HTML element of the webpages indicates whether it should be classified as main content or not, and whether it should be classified as template or not. In addition, the main menu of the webpages has also been labelled. The suite also incorporates scripts to automatize the benchmarking process. 

This suite has been developed as the result of a research project. We developed a new technique for content extraction \cite{InsST13} that was later adapted for template detection \cite{AlaIST13}. In the evaluation phase, our initial intention was to use a public benchmark suite. We first tried to use the CleanEval \cite{Baroni2008CleanevalAC} suite of content extraction benchmarks, because it has been widely used in the literature. Unfortunately, it is not prepared for template detection. Then, we contacted the authors of other techniques that had already evaluated their techniques. However, we could not use these benchmarks due to privacy (they belong to a company or project whose results were not shared), copyright (they were not publicly available) or unavailability (they had been lost). Finally, we decided to build or own benchmark suite and make it free and publicly available. 
Later, we developed a new menu detection technique \cite{DBLP:conf/sofsem/AlarteIS17}, so we decided to update the suite by labelling the nodes that represent the main menu of the website. Hence, we included 25 new benchmarks, for a total of 65. Then, we doubled the number of benchmarks, up to 130. Finally, the last year we have added 20 more benchmarks, therefore it is now formed from 150 websites.
The rest of this paper presents that benchmark suite.

%%%%%%%%%%%%%%%%%%%%%%%%%%%%%%%%%%%%%%%%%%%%%%%

\section{The TECO Benchmark Suite}\label{sec-TEBS}

TECO (TEmplate detection and COntent extraction benchmarks suite) was created as a benchmark suite specifically designed for template detection and content extraction. It can be used for testing and evaluation of these techniques, 
and it is formed from 150 real websites downloaded from Internet. 
We selected heterogenous websites such as blogs, companies, forums, personal websites, sports websites, newspapers, etc. Some of the websites are well known, like the BBC website or the Apple website, and others are less known like personal blogs or small companies websites.
Most of the benchmarks are English or Spanish websites, but we can also find websites in other languages, such as Italian, German, French, Portuguese, Dutch, Chinese, and Greek. 

The downloading of the webpages was done in some cases using the OS X software SiteSucker, and in other cases using the Linux command \texttt{wget}.

%\textbf{TODO: Â¿Incluir par\'ametros o configuraciones de los programas?. Incluido en el ejemplo.}\\
It is important to know how the websites were downloaded and stored, so that other researchers can increase the suite if it is needed. The following command downloads a website from the Linux terminal using the \texttt{wget} command:\\
\texttt{\$ wget --convert-links --no-clobber --random-wait -r 3 -p -E -e \\robots=off -U mozilla \textit{http://www.example.org}}\\
The meaning of the flags used is:
\begin{itemize}
\renewcommand{\labelitemi}{$\bullet$}
\item \texttt{--convert-links:} Converts links so they can work locally.
\item \texttt{--no-clobber:} Do not overwrite any existing file.
\item \texttt{--random-wait:} Random waits between downloads.
\item \texttt{-r 3:} Recursive downloading up to 3 levels of links.
\item \texttt{-p:} Downloads everything.
\item \texttt{-e robots=off:} Act as not being a robot.
\item \texttt{-E:} Get the right file extension.
\item \texttt{-U mozilla:} Identify as a Mozilla browser.
\end{itemize}

Each benchmark is composed of:
\begin{itemize} 
\item A principal webpage, called \emph{key page}. It is the target webpage from which the techniques should extract the main content or the template---note that it is not necessarily the main webpage of the website (e.g., index.html)---. 
\item A set of webpages that belong to the same website as the key page. This set contains all those webpages that are linked by the key page, and also the webpages linked by them.
\end{itemize}

\subsection{Producing the gold standard}
The suite comes with a gold standard that can be used as a reference to compare different techniques. 
The gold standard specifies for each key page what parts form the template. This is indicated in the own webpage by using HTML classes that indicate what elements are classified as \emph{notTemplate}. 
It has been produced manually by careful inspection of the websites and mixing the opinion of several people.   

In particular, once all the websites were downloaded (the key page and two levels of linked webpages in the same domain), four different engineers did the following independently:
\begin{itemize}
\item They manually explored the key page and the webpages accessible from it to decide what part of the webpage is the template and what part is the main content.
\item They printed the template and the main content of the webpage.
\end{itemize}

Then, the four engineers met and performed again these two actions but now all together sharing their individual opinions. Using the results of this agreement, each website was prepared for both, template extraction and content detection. 
On the one hand, all elements from the key page not belonging to the template were included in a HTML class called \textit{TECO\_notTemplate}. This way, a template extraction tool can automatically compare its output with the nodes not belonging to the \textit{TECO\_notTemplate} class.
On the other hand, all elements belonging to the main content were included in an HTML class called \textit{TECO\_mainContent}. Therefore, a content extraction tool can easily compare its output with the nodes belonging to that class.
In addition, the node that represents the main menu of the webpage was included in an HTML class called \textit{TECO\_mainMenu}. Consequently, a menu extraction tool can compare its output with the nodes belonging to that class.

\subsection{Benchmark details}

A classification of the benchmarks is important and useful depending on the application and technique that is being fed with them. We provide different classifications according to the purpose and properties of the benchmarks.  
First, all benchmarks have been classified into five groups: 
\smallskip

Companies / Shops, \hspace{.5cm} Forums / Social, \hspace{.5cm} Personal websites / Blogs, 

Media / Communication, \hspace{.5cm} Institutions / Associations.

\smallskip
\noindent Table \ref{tbl-URLs} shows this classification together with the URLs from which we extracted the benchmarks.

%Primera parte de la tabla
\begin{table}[htp!]
\begin{center}
\caption{Sources of the benchmarks}
\scalebox{0.8}{
\begin{tabular}{|ll|}
\hline
\multicolumn{1}{|l|}{\textbf{Website type}} & \textbf{Original URL of the webpage}                                                                  \\ \hline
Forums / Social             & es.sharelatex.com/learn/Uploading\_a\_project                                                                        \\
                            & github.com/DawidStankiewicz/forum                                                                                  \\
                            & en.citizendium.org                                                                                        \\
                            & www.filmaffinity.com/es/                                                                                    \\
                            & www.meneame.net/faq-es/                                                                                          \\
                            & www.accountkiller.com/en/delete-activision-account                                                              \\
                            & study.com/learn/science-questions-and-answers.html                                                                   \\
                            & c.mi.com/it/                                                                                               \\
                            & alumni.harvard.edu/help/message-board/                                                                           \\
                            & www.spacetimestudios.com/forumdisplay.php?29-Websites-and-Forum-Discussion
\\                            
							& www.gimpforum.de
\\							
							& www.emaildiscussions.com
\\							
							& forums.debian.net/viewforum.php?f=5
\\							
							& forums.mozillazine.org/viewforum.php?f=23
\\							
							& forums.tomsguide.com/forums/laptop-general-discussion.15/
\\							
							& forums.mysql.com/list.php?21
\\
							& lawstudents.ca/forums.html
\\							
							& www.japanesepod101.com/forum/viewforum.php?f=26
\\							                          
                            & forum.skyscraperpage.com                                                                                  \\
							& forums.opera.com
\\
							& forums.linuxmint.com/viewforum.php?f=72
\\
                            & frances.forosactivos.net                                                                                  \\
							& www.wysiwygwebbuilder.com/forum/viewforum.php?f=10
\\							                            
                            & www.3dprintforums.com                                                                                         \\
                            & www.strangehorizons.com/2004/20040906/greenglass-f.shtml                                                        
\\
							& communities.apple.com/es/community/mac\_os/os\_x\_el\_capitan.html
\\
							& www.sloweurope.com/community/
\\
							& community.ricksteves.com/travel-forum/spain.html
\\
							& hackercombat.com/forum/
\\
							& www.scbwi.org/boards/index.php?board=62.0
\\
							
Personal / Blogs            & www.cocinaconmarta.com/2015/04/empanadillas-chinas-de-gambas-y-verduras.html                                         \\
                            & www.trendencias.com                                                                                                  \\
                            & googleblog.blogspot.com.es                                                                                           \\
                            & www.robyncarr.com/qa/                                                                                            \\
                            & users.dsic.upv.es/$\sim$jsilva/wwv2013/index2.html                                                                   \\
                            & www.folj.com/puzzles/difficult-logic-problems.htm                                                                    \\
                            & oneminutelist.com/16-browser-alternatives-to-desktop-programs/                                             \\
							& artsonline.uwaterloo.ca/jburbidg/index.html                                                                                             \\
							& benjamincongdon.me/blog.html                                                                                             \\
							& michael.tsikerdekis.com                                                                                             \\
							& www.beeorganisee.com/reprendre-en-main-le-nettoyage/                                                                                             \\
							& www.danielgrindrod.com/about.html                                                                                             \\
							& ofdollarsanddata.com                                                                                             \\
							& blog.mint.com/updates/enter-our-newdecadenewyou-meme-sweepstakes-for-a-chance-to-win-5000/                                                                                             \\
							& elainesir.com/best-korean-beauty-blogs-bloggers-follow/                                                                                             \\
							& www.vindame.com.br/semana-riesling/uva-riesling/                                                                                             \\
							& www.rosamontero.es/obra-rosa-montero.html                                                                                             \\
							& www.almezzer.com/libros/literatura-infantil/a-partir-de-4-anos/                                                                                             \\
                            & markahall.blogspot.com.es                                                                                            \\
							& johnboyne.com/about/                                                                                             \\
                            & users.dsic.upv.es/$\sim$dinsa/en/                                                                          \\
							& johngardnerathome.info                                                                                             \\                                                  
                            & www.annmalaspina.com                                                                                      \\
                            & foodsense.is/a-list.html                                                                                             \\
							& sites.google.com/a/ciencias.unam.mx/pagina-ana-meda/
\\
							& whatever.scalzi.com/about/interviews-appearances-articles-and-etc/
\\
                            & www.javiercelaya.es                                                                                       \\
                            & diarium.usal.es/lguich/pagina-personal-de-luis-arturo-guichard/                                                       \\
							& www.jameslovelock.org/scientific-papers/
\\
							& www.cipri.info    
\\
\hline
\end{tabular}
\label{tbl-URLs}
}
\end{center}
\end{table}%

%Segunda parte tabla
\begin{table}[htp!]
%\caption{Sources of the benchmarks}
\begin{center}
\scalebox{0.8}{
\begin{tabular}{|ll|}
\hline
\multicolumn{1}{|l|}{\textbf{Website type}} & \textbf{Original URL of the webpage}                                                                  \\ \hline
Companies / Shops           & today.java.net/pub/a/today/2004/07/06/3ddesktop.html                                                                 \\
                            & clotheshor.se                                                                                             \\
                            & www.raspberrypi.org/resources/teach/                                                                       \\
                            & doodle.com/online-calendar/                                                                                      \\
                            & www.newprosoft.com/web-content-extractor.htm                                                                         \\
                            & worryfreelabs.com/about/                                                                                       \\
                            & www.intelligencetest.com                                                                                   \\
                            & www.ikea.com/gb/                                                                                              \\
                            & www.nubbeo.com.ar
\\                            
							& www.mulberry.com/es/shop/sale/sale-mens-accessories.html
\\							
							& www.tous.com/es-es/novedades/relojes/c/59.html
\\							
							& preferenceweb.com/collections/all-sneakers.html
\\
							& www.trekbikes.com/us/en\_US/bikes/mountain-bikes/electric-mountain-bikes/c/B512/
\\							
							& addons.prestashop.com/es/2-modulos.html
\\							
							& us.pandora.net/en/charm-bracelets/pandora-moments/pandora-moments-bracelets/
\\							
							& kawaiipenshop.com
\\
							& www.vam.ac.uk/shop/lindsay-philip-butterfield-blue-flower-silk-scarf.html
\\							
							& shop.fendt.com/kids-toys/clothing/shirts.html
\\							
							& www.euroholds.com/it/29-prese-arrampicata.html
\\							                         
                            & www.emmaclothes.com                                                                                       \\
                            & www.arduino.cc/en/Main/Software/                                                                                 \\
							& naranjascarcaixent.com/tienda.html
\\							
							& www.technicalbookstoreonline.com/new-arrivals.php
\\
							& www.floridarealestatecollege.com
\\
							& www.basf.com/nl/nl/who-we-are/BASF-in-Nederland.html
\\
							& www.mcphersonoil.com
\\
							& www.thirteenhou.com/menu.php
\\
							& wwww.embalajesterra.com/precintadoras-manuales-168
\\
							& www.crypto.ch/en/about
\\
							& www.shopbookshop.com
\\							
Media / Communication       & edition.cnn.com                                                                                           \\
                            & www.neoteo.com/star-wars-the-force-awakens-el-regreso-de-viejos-personajes/                                          \\
                            & riotimesonline.com                                                                                        \\
                            & www.turfparadise.com                                                                                      \\
                            & www.cleanclothes.org                                                                                      \\
                            & www.afp.com/es/contact/                                                                                          \\
                            & www.history.com                                                                                           \\
                            & detroit.cbslocal.com/2018/12/04/high-school-newspaper-suspended-after-publishing-disruptive-investigation/ \\
                            & www.rocklists.com/91x-1983.html                                                                                      \\
                            & www.lashorasperdidas.com                                                                                  \\
                            & www.journalism.org/2014/03/13/social-search-direct/
\\                           
							& www.socialmediatoday.com/news/facebook-adds-new-features-for-instant-articles-including-links-to-more-pu/	\\	&	569786/
\\							
							& www.diariodeburgos.es/Noticia/Z1C5D6DE9-D1E6-B03A-61236AF21520B8B2/202002/Un-programa-verde-	\\	&	dedicado-a-Felix-Rodriguez-de-la-Fuente.html
\\									
							& wordofmouthmendo.com/word-of-mouth-stories/2018/5/31/travellers-fare.html
\\	
							& www.usine-digitale.fr/article/la-start-up-americaine-clearview-ai-illustre-deja-les-derives-de-la-reconnaissance	\\	&	-faciale.N921119.html
\\						
							& 1015fm.com.au/2020/02/steve-mickenbecker-interest-rates-on-hold-2020-02-07/
\\							
							& www.dw.com/de/lebron-james-vom-pflegekind-zum-basketball-superstar/a-52088565.html
\\							
							& www.theday.com/movies--tv/20200203/super-bowl-ads-dialed-up-fun-as-antidote-to-politics.html
\\							
							& nltimes.nl/2019/12/16/chocolate-spread-babies-wins-misleading-product-award.html
\\							                         
                            & www.bbc.co.uk/news/                                                                                        \\
                            & techcrunch.com/gadgets/                                                                                               \\
							& biztechmagazine.com/article/2019/12/why-byod-makes-endpoint-security-crucial-small-businesses.html
\\							
							& www.eeo.com.cn/2022/0506/533366.shtml
\\							
							& www.wishtv.com/news/flu-is-widespread-across-the-us/
\\							
							& news.mit.edu/2021/grand-decoding-data-0909.html
\\
							& asia.nikkei.com/Spotlight/Sharing-Economy/New-Tokyo-homes-ditch-parking-spaces-but-offer-car-sharing
\\
							& www.rcnky.com/articles/2021/09/12/ft-mitchell-reflects-life-age-104.html
\\                      
                            & news.discovery.com/tech/robotics/artificial-intelligences-hawkings-fears-stir-debate-141206.htm
\\
							& www.kathimerini.gr/society/561833251/koronoios-arsi-metron-i-megali-prova-kanonikotitas-enopsei-toy-kalokairioy/
\\
							& news.un.org/en/content/navigate-news
\\		
\hline
\end{tabular}
}
\end{center}
%\label{tbl-URLs}
\end{table}%

%Tercera parte tabla
\begin{table}[t!]
%\caption{Sources of the benchmarks}
\begin{center}
\scalebox{0.8}{
\begin{tabular}{|ll|}
\hline
\multicolumn{1}{|l|}{\textbf{Website type}} & \textbf{Original URL of the webpage}                                                                  \\ \hline							
Institutions / Associations & web.mit.edu/institute-events/visitor/                                                                                 \\
                            & www.museodelprado.es                                                                                      \\
                            & www.u-tokyo.ac.jp/en/about/history.html                                                                              \\
                            & www.savethechildren.net/what-we-do/our-humanitarian-work/                                                        \\
                            & college.harvard.edu/financial-aid/                                                                               \\
							& www.linuxfoundation.org/about/                                                                                 \\
                            & clinicaltrials.gov/ct2/search/index/                                                                       \\
                            & cordis.europa.eu/fp7/ict/fire.html                                                                                   \\
                            & parents.berkeley.edu/advice/babies/laundry.html
\\ 
							& www.mit.edu/campus-life
\\
							& cpoepalencia.es/federaciones-y-asociaciones-confederadas-asociaciones/
\\							
							& www.icann.org/history.html
\\							
							& www.gip-jci-justice.fr/en/about-us/support-council/
\\
							& www.einstein.yu.edu/leadership/
\\							
							& www.americanacademy.de/about/
\\							
							& www.mensa.es/cms/pages/\%C2\%BFqu\%C3\%A9-es-mensa.html
\\							
							& www.bcrf.org/breast-cancer-research.html
\\							
							& www.ielts.org/what-is-ielts/ielts-introduction.html
\\							
							& fr.unesco.org/about-us/introducing-unesco.html
\\							
							& www.ccbe.eu/about/who-we-are/
\\							
							& www.fraud.org/get\_involved.html
\\							
                            & www.jdi.org.za                                                                                            \\
							& www.premiere-urgence.org/qui-sommes-nous/
\\
							& www.indiangaming.org
\\
							& hispalinux.es/QuienesSomos
\\
							& www.gktw.org/about/
\\
							& www.apnic.net/about-apnic/organization/vision-mission-objectives/
\\
                            & www.unicef.org/where-we-work.html                                                                                    \\
                            & www.klimabuendnis.org/home.html
\\                     				
                            & www.isoc-es.org                                                                                                      \\
\hline
\end{tabular}
}
\end{center}
%\label{tbl-URLs}
\end{table}%

Table \ref{tbl-bench_prop} shows some properties of the benchmarks. 
Here, 
column \textbf{Nodes} indicates the total number of DOM nodes in the key page, 
column \textbf{T. Nodes} shows the number of DOM nodes that belong to the template and 
column \textbf{M.C. Nodes} refers to the number of DOM nodes that belong to the main content. Note that these values are not necessarily complementary. I.e., many webpages contain DOM nodes that do not belong to the main content nor to the template. For instance, usually webpages contain irrelevant (non-main content) information that only appears in that webpage, such as submenus, blocks from social networks, sliders, etc.

%Primera parte de la tabla
\begin{table}[htp!]
\caption{Benchmark properties}
\begin{center}
\scalebox{0.8}{
\begin{tabular}{| l | l | r | r | r | r |}
\hline
\textbf{Id} & \textbf{Benchmark's domain}  & \textbf{Nodes}  & \textbf{T. Nodes}  & \textbf{M.C. Nodes} & \textbf{Menu Links} \\ \hline
1 & web.mit.edu & 424 & 252 & 141 & 9 \\
2 & www.museodelprado.es & 639 & 148 & 168 & 7 \\
3 & www.u-tokyo.ac.jp & 614 & 499 & 97 & 30 \\
4 & www.savethechildren.net & 763 & 690 & 54 & 21 \\
5 & college.harvard.edu & 1098 & 669 & 397 & 5 \\
6 & www.linuxfoundation.org & 597 & 534 & 38 & 118 \\
7 & clinicaltrials.gov & 545 & 424 & 101 & 37 \\
8 & cordis.europa.eu & 980 & 335 & 164 & 19 \\
9 & parents.berkeley.edu & 287 & 99 & 180 & 8 \\
10 & www.mit.edu & 1290 & 472 & 809 & 13 \\
11 & cpoepalencia.es & 719 & 644 & 73 & 51 \\
12 & www.icann.org & 492 & 397 & 90 & 46 \\
13 & www.gip-jci-justice.fr & 887 & 680 & 137 & 28 \\
14 & www.einstein.yu.edu & 1168 & 815 & 187 & 29 \\
15 & www.americanacademy.de & 746 & 670 & 14 & 37 \\
16 & www.mensa.es & 422 & 354 & 37 & 10 \\
17 & www.bcrf.org & 917 & 587 & 294 & 6 \\
18 & www.ielts.org & 761 & 605 & 150 & 43 \\
19 & fr.unesco.org & 957 & 615 & 308 & 73 \\
20 & www.ccbe.eu & 1003 & 783 & 177 & 33 \\
21 & www.fraud.org & 558 & 321 & 61 & 16 \\
22 & www.jdi.org.za & 661 & 401 & 199 & 10 \\
23 & www.premiere-urgence.org & 502 & 457 & 32 & 30 \\
24 & www.indiangaming.org & 594 & 209 & 148 & 7 \\
25 & hispalinux.es & 515 & 347 & 144 & 32 \\
26 & www.gktw.org & 793 & 646 & 130 & 8 \\
27 & www.apnic.net & 650 & 461 & 79 & 59 \\
28 & www.unicef.org & 1057 & 671 & 381 & 4 \\
29 & www.klimabuendnis.org & 892 & 536 & 134 & 75 \\
30 & www.isoc-es.org & 279 & 171 & 56 & 17 \\
31 & edition.cnn.com & 3980 & 192 & 877 & 15 \\
32 & www.neoteo.com & 1051 & 636 & 388 & 18 \\
33 & riotimesonline.com & 2115 & 1094 & 743 & 23 \\
34 & www.turfparadise.com & 1072 & 838 & 205 & 98 \\
35 & www.cleanclothes.org & 1358 & 266 & 932 & 7 \\
36 & www.afp.com & 1208 & 404 & 789 & 16 \\
37 & www.history.com & 1324 & 673 & 260 & 12 \\
38 & detroit.cbslocal.com & 1261 & 1004 & 96 & 65 \\
39 & www.rocklists.com & 783 & 533 & 184 & 6 \\
40 & www.lashorasperdidas.com & 1924 & 554 & 683 & 12 \\
41 & www.journalism.org & 830 & 459 & 86 & 10 \\
42 & www.socialmediatoday.com & 1288 & 666 & 149 & 8 \\
43 & www.diariodeburgos.es & 606 & 384 & 69 & 9 \\
44 & wordofmouthmendo.com & 916 & 668 & 26 & 26 \\
45 & www.usine-digitale.fr & 994 & 259 & 124 & 18 \\
46 & 1015fm.com.au & 1041 & 835 & 65 & 28 \\
47 & www.dw.com & 2593 & 1596 & 470 & 135 \\
48 & www.theday.com & 2147 & 933 & 456 & 86 \\
49 & nltimes.nl & 588 & 115 & 164 & 10 \\
50 & www.bbc.co.uk & 3029 & 573 & 1195 & 22 \\
51 & techcrunch.com & 2612 & 1891 & 586 & 35 \\
52 & biztechmagazine.com & 1950 & 1057 & 454 & 97 \\
53 & www.eeo.com.cn & 936 & 676 & 119 & 11 \\
54 & www.wishtv.com & 2380 & 1993 & 343 & 77 \\
55 & news.mit.edu & 2122 & 1045 & 128 & 8 \\
56 & asia.nikkei.com & 886 & 671 & 116 & 42 \\
57 & www.rcnky.com & 1771 & 1435 & 112 & 17 \\
58 & news.discovery.com & 2926 & 1209 & 791 & 68 \\
59 & www.kathimerini.gr & 1897 & 1606 & 117 & 83 \\
60 & news.un.org & 1809 & 1258 & 59 & 42 \\
61 & es.sharelatex.com & 1100 & 877 & 214 & 6 \\
62 & github.com & 1242 & 453 & 783 & 5 \\
63 & en.citizendium.org & 1092 & 415 & 633 & 35 \\
64 & www.filmaffinity.com & 1337 & 352 & 972 & 32 \\
65 & www.meneame.net & 769 & 207 & 423 & 11 \\
\hline
\end{tabular}
}
\end{center}
\label{tbl-bench_prop}
\end{table}

%Segunda parte de la tabla
\begin{table}[htp!]
%\caption{Benchmark properties}
\begin{center}
\scalebox{0.8}{
\begin{tabular}{| l | l | r | r | r | r |}
\hline
\textbf{Id} & \textbf{Benchmark's domain}  & \textbf{Nodes}  & \textbf{T. Nodes}  & \textbf{M.C. Nodes} & \textbf{Menu Links} \\ \hline
66 & www.accountkiller.com & 510 & 222 & 273 & 8 \\
67 & study.com & 7328 & 1897 & 5433 & 74 \\
68 & c.mi.com & 3506 & 2949 & 541 & 37 \\
69 & alumni.harvard.edu & 2026 & 1785 & 219 & 40 \\
70 & www.spacetimestudios.com & 5049 & 1387 & 3500 & 47 \\
71 & www.gimpforum.de & 2058 & 457 & 1300 & 6 \\
72 & www.emaildiscussions.com & 1129 & 239 & 674 & 7 \\
73 & forums.debian.net & 2766 & 150 & 2327 & 7 \\
74 & forums.mozillazine.org & 2023 & 235 & 1411 & 4 \\
75 & forums.tomsguide.com & 7911 & 992 & 5064 & 22 \\
76 & forums.mysql.com & 4493 & 430 & 3950 & 10 \\
77 & lawstudents.ca & 3563 & 949 & 1935 & 11 \\
78 & www.japanesepod101.com & 1574 & 924 & 434 & 36 \\
79 & forum.skyscraperpage.com & 3410 & 146 & 1676 & 6 \\
80 & forums.opera.com & 1456 & 617 & 829 & 7 \\
81 & forums.linuxmint.com & 5079 & 327 & 3872 & 7 \\
82 & frances.forosactivos.net & 814 & 318 & 495 & 9 \\
83 & www.wysiwygwebbuilder.com & 3941 & 739 & 3201 & 7 \\
84 & www.3dprintforums.com & 1125 & 312 & 748 & 8 \\
85 & www.strangehorizons.com & 643 & 149 & 406 & 23 \\
86 & communities.apple.com & 3144 & 375 & 1306 & 10 \\
87 & www.sloweurope.com & 4208 & 526 & 2789 & 29 \\
88 & community.ricksteves.com & 2061 & 386 & 1177 & 9 \\
89 & hackercombat.com & 1828 & 828 & 698 & 6 \\
90 & www.scbwi.org & 892 & 219 & 506 & 6 \\
91 & www.cocinaconmarta.com & 4154 & 3404 & 307 & 9 \\
92 & www.trendencias.com & 2503 & 1139 & 1040 & 7 \\
93 & googleblog.blogspot.com.es & 5096 & 3574 & 1494 & 343 \\
94 & www.robyncarr.com & 292 & 92 & 200 & 4 \\
95 & users.dsic.upv.es & 207 & 170 & 34 & 14 \\
96 & www.folj.com & 567 & 176 & 384 & 4 \\
97 & oneminutelist.com & 503 & 276 & 211 & 5 \\
98 & artsonline.uwaterloo.ca & 413 & 164 & 240 & 4 \\
99 & benjamincongdon.me & 329 & 55 & 274 & 4 \\
100 & michael.tsikerdekis.com & 577 & 124 & 95 & 7 \\
101 & www.beeorganisee.com & 840 & 494 & 303 & 18 \\
102 & www.danielgrindrod.com & 424 & 395 & 29 & 4 \\
103 & ofdollarsanddata.com & 1075 & 365 & 651 & 5 \\
104 & blog.mint.com & 872 & 442 & 129 & 44 \\
105 & elainesir.com & 1383 & 523 & 636 & 26 \\
106 & www.vindame.com.br & 667 & 546 & 104 & 20 \\
107 & www.rosamontero.es & 808 & 89 & 717 & 9 \\
108 & www.almezzer.com & 1121 & 516 & 416 & 23 \\
109 & markahall.blogspot.com.es & 3144 & 697 & 2437 & 22 \\
110 & johnboyne.com & 690 & 214 & 185 & 8 \\
111 & users.dsic.upv.es & 243 & 75 & 160 & 5 \\
112 & johngardnerathome.info & 397 & 176 & 188 & 21 \\
113 & www.annmalaspina.com & 403 & 190 & 84 & 8 \\
114 & foodsense.is & 339 & 104 & 192 & 5 \\
115 & sites.google.com & 405 & 320 & 85 & 32 \\
116 & whatever.scalzi.com & 1693 & 1434 & 243 & 11 \\
117 & www.javiercelaya.es & 763 & 682 & 57 & 12 \\
118 & diarium.usal.es & 625 & 82 & 524 & 3 \\
119 & www.jameslovelock.org & 679 & 465 & 174 & 20 \\
120 & www.cipri.info & 955 & 387 & 556 & 27 \\
121 & today.java.net & 733 & 342 & 354 & 6 \\
122 & clotheshor.se & 465 & 232 & 228 & 8 \\
123 & www.raspberrypi.org & 398 & 143 & 209 & 14 \\
124 & doodle.com & 580 & 491 & 82 & 5 \\
125 & www.newprosoft.com & 833 & 151 & 679 & 6 \\
126 & worryfreelabs.com & 514 & 321 & 190 & 7 \\
127 & www.intelligencetest.com & 595 & 323 & 263 & 18 \\
128 & www.ikea.com & 1556 & 407 & 985 & 10 \\
129 & www.nubbeo.com.ar & 1642 & 605 & 975 & 7 \\
130 & www.mulberry.com & 8506 & 3943 & 4203 & 152 \\
\hline
\end{tabular}
}
\end{center}
%\label{tbl-bench_prop}
\end{table}

%Tercera parte de la tabla
\begin{table}[t!]
\begin{center}
\scalebox{0.8}{
\begin{tabular}{| l | l | r | r | r | r |}
\hline
\textbf{Id} & \textbf{Benchmark's domain}  & \textbf{Nodes}  & \textbf{T. Nodes}  & \textbf{M.C. Nodes} & \textbf{Menu Links} \\ \hline
131 & www.tous.com & 5109 & 3010 & 1056 & 216 \\
132 & preferenceweb.com & 2279 & 725 & 1322 & 17 \\
133 & www.trekbikes.com & 5698 & 1924 & 3760 & 15 \\
134 & addons.prestashop.com & 8062 & 2333 & 3382 & 5 \\
135 & us.pandora.net & 6375 & 2011 & 3376 & 142 \\
136 & kawaiipenshop.com & 1237 & 821 & 403 & 26 \\
137 & www.vam.ac.uk & 1595 & 1186 & 392 & 61 \\
138 & shop.fendt.com & 2278 & 1433 & 640 & 55 \\
139 & www.euroholds.com & 4923 & 843 & 3490 & 89 \\
140 & www.emmaclothes.com & 1088 & 374 & 705 & 8 \\
141 & www.arduino.cc & 854 & 490 & 336 & 26 \\
142 & naranjascarcaixent.com & 321 & 172 & 141 & 6 \\
143 & www.technicalbookstoreonline.com & 2971 & 391 & 2002 & 12 \\
144 & www.floridarealestatecollege.com & 1069 & 556 & 65 & 29 \\
145 & www.basf.com & 845 & 776 & 62 & 12 \\
146 & www.mcphersonoil.com & 891 & 628 & 225 & 34 \\
147 & www.thirteenhou.com & 1226 & 137 & 1073 & 4 \\
148 & www.embalajesterra.com & 2360 & 1694 & 470 & 106 \\
149 & www.crypto.ch & 346 & 249 & 68 & 3 \\
150 & www.shopbookshop.com & 1743 & 1314 & 387 & 9          \\ \hline
\end{tabular}
}
\end{center}
%\label{tbl-bench_prop}
\end{table}

\subsection{Guidelines for using the suite}
\subsubsection{Downloading and configuring the suite\\}

TECO is freely distributed and can be downloaded from the URL: 
\begin{center}
http://www.dsic.upv.es/$\sim$jsilva/retrieval/teco
\end{center}

After downloading the suite, a directory that contains 150 folders, one for each website, is created. 
Table \ref{tbl-paths} shows the path to the key page of each benchmark.

%TECO is also distributed with a Firefox add-on to check the benchmarks. This extension can also be downloaded from the TECO website. The extension is a zip file which includes all the necessary files. This zip file has to be uncompressed in order to be configured. Once uncompressed, it will create two directories called \textit{chrome} and \textit{defaults}, and two filles called \textit{install.rdf} and \textit{chrome.manifest}. All the configuration can be modified by editing the \textit{benchmarks.js} file which is located in the \textit{/chrome/content} directory.
%
%First of all, the benchmarks path needs to be configured, it can be done through two variables which are:
%\begin{itemize}
%\item \textbf{initialPath:} It is the local path. \textit{Example: /home/}
%\item \textbf{pathBenchmarks:} The common path that all benchmarks share. \textit{Example: file:///home/asus/}
%\end{itemize}

%Primera parte de la tabla
\begin{table}[htp!]
\begin{center}
\caption{Path to the key page of each benchmark}
%\begin{small}
\scalebox{0.8}{
\begin{tabular}{| l | p{1.1\linewidth} |}
\hline
\textbf{Id} & \textbf{Path to the key page}                                                                         \\ \hline
1 & web.mit.edu/institute-events/visitor \\
2 & www.museodelprado.es/index.html \\
3 & www.u-tokyo.ac.jp/en/about/history.html \\
4 & www.savethechildren.net/what-we-do/our-humanitarian-work.html \\
5 & college.harvard.edu/financial-aid.html \\
6 & www.linuxfoundation.org/about.1.html \\
7 & clinicaltrials.gov/ct2/search/index/index.html \\
8 & cordis.europa.eu/fp7/ict/fire.html \\
9 & parents.berkeley.edu/advice/babies/laundry.html \\
10 & www.mit.edu/campus-life.1.html \\
11 & cpoepalencia.es/federaciones-y-asociaciones-confederadas-asociaciones/index.html \\
12 & www.icann.org/history.html \\
13 & www.gip-jci-justice.fr/en/about-us/support-council/index.html \\
14 & www.einstein.yu.edu/leadership/index.html \\
15 & www.americanacademy.de/about/index.html \\
16 & www.mensa.es/cms/pages/\%C2\%BFqu\%C3\%A9-es-mensa.html \\
17 & www.bcrf.org/breast-cancer-research.html \\
18 & www.ielts.org/what-is-ielts/ielts-introduction.html \\
19 & fr.unesco.org/about-us/introducing-unesco.html \\
20 & www.ccbe.eu/about/who-we-are/index.html \\
21 & www.fraud.org/get\_involved.html \\
22 & www.jdi.org.za/index.html \\
23 & www.premiere-urgence.org/qui-sommes-nous/index.html \\
24 & www.indiangaming.org/index.html \\
25 & hispalinux.es/QuienesSomos.html \\
26 & www.gktw.org/about/index.html \\
27 & www.apnic.net/about-apnic/organization/vision-mission-objectives/index.html \\
28 & www.unicef.org/where-we-work.html \\
29 & www.klimabuendnis.org/home.html \\
30 & www.isoc-es.org \\
31 & edition.cnn.com/index.html \\
32 & www.neoteo.com/star-wars-the-force-awakens-el-regreso-de-viejos-personajes/ \\
33 & riotimesonline.com/index.html \\
34 & www.turfparadise.com/index.html \\
35 & www.cleanclothes.org/index.html \\
36 & www.afp.com/es/contact.html \\
37 & www.history.com/index.html \\
38 & detroit.cbslocal.com/2018/12/04/high-school-newspaper-suspended-after-publishing-disruptive-investigation/index.html \\
39 & www.rocklists.com/91x-1983.html \\
40 & www.lashorasperdidas.com/index.html \\
41 & www.journalism.org/2014/03/13/social-search-direct/index.html \\
42 & www.socialmediatoday.com/news/facebook-adds-new-features-for-instant-articles-including-links-to-more-pu/569786/index.html \\
43 & www.diariodeburgos.es/Noticia/Z1C5D6DE9-D1E6-B03A-61236AF21520B8B2/202002/Un-programa-verde-dedicado-a-Felix-Rodriguez-de-la-Fuente.html \\
44 & wordofmouthmendo.com/word-of-mouth-stories/2018/5/31/travellers-fare.html \\
45 & www.usine-digitale.fr/article/la-start-up-americaine-clearview-ai-illustre-deja-les-derives-de-la-reconnaissance-faciale.N921119.html \\
46 & 1015fm.com.au/2020/02/steve-mickenbecker-interest-rates-on-hold-2020-02-07/index.html \\
47 & www.dw.com/de/lebron-james-vom-pflegekind-zum-basketball-superstar/a-52088565.html \\
48 & www.theday.com/movies--tv/20200203/super-bowl-ads-dialed-up-fun-as-antidote-to-politics.html \\
49 & nltimes.nl/2019/12/16/chocolate-spread-babies-wins-misleading-product-award.html \\
50 & www.bbc.co.uk/news/index.html \\
51 & techcrunch.com/gadgets \\
52 & biztechmagazine.com/article/2019/12/why-byod-makes-endpoint-security-crucial-small-businesses.html \\
53 & www.eeo.com.cn/2022/0506/533366.shtml.html \\
54 & www.wishtv.com/news/flu-is-widespread-across-the-us/index.html \\
55 & news.mit.edu/2021/grand-decoding-data-0909.html \\
56 & asia.nikkei.com/Spotlight/Sharing-Economy/New-Tokyo-homes-ditch-parking-spaces-but-offer-car-sharing.html \\
57 & www.rcnky.com/articles/2021/09/12/ft-mitchell-reflects-life-age-104.html \\
\hline
\end{tabular}
\label{tbl-paths}
}
\end{center}
\end{table}

%Segunda parte de la tabla
\begin{table}[htp!]
%\caption{Path to the key page of each benchmark}
\begin{center}
%\begin{small}
\scalebox{0.8}{
\begin{tabular}{| l | p{1.1\linewidth} |}
\hline
\textbf{Id} & \textbf{Path to the key page}                                                                         \\ \hline
58 & news.discovery.com/tech/robotics/artificial-intelligences-hawkings-fears-stir-debate-141206.htm \\
59 & www.kathimerini.gr/society/561833251/koronoios-arsi-metron-i-megali-prova-kanonikotitas-enopsei-toy-kalokairioy/index.html \\
60 & news.un.org/en/content/navigate-news.html \\
61 & es.sharelatex.com/learn/Uploading\_a\_project \\
62 & github.com/DawidStankiewicz/forum.1 \\
63 & en.citizendium.org/index.html \\
64 & www.filmaffinity.com/es/main.html \\
65 & www.meneame.net/faq-es.html \\
66 & www.accountkiller.com/en/delete-activision-account.html \\
67 & study.com/learn/science-questions-and-answers.html \\
68 & c.mi.com/it/index.html \\
69 & alumni.harvard.edu/help/message-board.html \\
70 & www.spacetimestudios.com/forumdisplay.php\%3f29-Websites-and-Forum-Discussion.html \\
71 & www.gimpforum.de/index.html \\
72 & www.emaildiscussions.com/index.html \\
73 & forums.debian.net/viewforum.php\%3ff\%3d5.html \\
74 & forums.mozillazine.org/viewforum.php\%3ff\%3d23.html \\
75 & forums.tomsguide.com/forums/laptop-general-discussion.15/index.html \\
76 & forums.mysql.com/list.php\%3f21.html \\
77 & lawstudents.ca/forums.html \\
78 & www.japanesepod101.com/forum/viewforum.php\%3ff\%3d26.html \\
79 & forum.skyscraperpage.com/index.html \\
80 & forums.opera.com/index.html \\
81 & forums.linuxmint.com/viewforum.php\%3ff\%3d72.html \\
82 & frances.forosactivos.net/index.html \\
83 & www.wysiwygwebbuilder.com/forum/viewforum.php\%3ff\%3d10.html \\
84 & www.3dprintforums.com/index.html \\
85 & www.strangehorizons.com/2004/20040906/greenglass-f.shtml.html \\
86 & communities.apple.com/es/community/mac\_os/os\_x\_el\_capitan.html \\
87 & www.sloweurope.com/community/index.html \\
88 & community.ricksteves.com/travel-forum/spain.html \\
89 & hackercombat.com/forum/index.html \\
90 & www.scbwi.org/boards/index.php\%3fboard\%3d62.0.html \\
91 & www.cocinaconmarta.com/2015/04/empanadillas-chinas-de-gambas-y-verduras.html \\
92 & www.trendencias.com \\
93 & googleblog.blogspot.com.es \\
94 & www.robyncarr.com/qa.html \\
95 & users.dsic.upv.es/$\sim$jsilva/wwv2013/index2.html \\
96 & www.folj.com/puzzles/difficult-logic-problems.htm \\
97 & oneminutelist.com/16-browser-alternatives-to-desktop-programs/index.html \\
98 & artsonline.uwaterloo.ca/jburbidg/index.html \\
99 & benjamincongdon.me/blog.html \\
100 & michael.tsikerdekis.com/index.html \\
101 & www.beeorganisee.com/reprendre-en-main-le-nettoyage/index.html \\
102 & www.danielgrindrod.com/about.html \\
103 & ofdollarsanddata.com/index.html \\
104 & blog.mint.com/updates/enter-our-newdecadenewyou-meme-sweepstakes-for-a-chance-to-win-5000/index.html \\
105 & elainesir.com/best-korean-beauty-blogs-bloggers-follow/index.html \\
106 & www.vindame.com.br/semana-riesling/uva-riesling/index.html \\
107 & www.rosamontero.es/obra-rosa-montero.html \\
108 & www.almezzer.com/libros/literatura-infantil/a-partir-de-4-anos/index.html \\
109 & markahall.blogspot.com.es \\
110 & johnboyne.com/about/index.html \\
111 & users.dsic.upv.es/$\sim$dinsa/en/index.html \\
112 & johngardnerathome.info/index.htm \\
113 & www.annmalaspina.com/index.html \\
114 & foodsense.is/a-list.html \\
115 & sites.google.com/a/ciencias.unam.mx/pagina-ana-meda/index.html \\
116 & whatever.scalzi.com/about/interviews-appearances-articles-and-etc/index.html \\
117 & www.javiercelaya.es/index.html \\
118 & diarium.usal.es/lguich/pagina-personal-de-luis-arturo-guichard \\
119 & www.jameslovelock.org/scientific-papers/index.html \\
120 & www.cipri.info/index.html \\
\hline
\end{tabular}
}
\end{center}
%\label{tbl-paths}
\end{table}

%Tercera parte de la tabla
\begin{table}[t!]
%\caption{Path to the key page of each benchmark}
\begin{center}
%\begin{small}
\scalebox{0.8}{
\begin{tabular}{| l | p{1.1\linewidth} |}
\hline
\textbf{Id} & \textbf{Path to the key page}                                                                         \\ \hline
121 & today.java.net/pub/a/today/2004/07/06/3ddesktop.html \\
122 & clotheshor.se/index.html \\
123 & www.raspberrypi.org/resources/teach/index.html \\
124 & doodle.com/online-calendar.html \\
125 & www.newprosoft.com/web-content-extractor.htm \\
126 & worryfreelabs.com/about.1.html \\
127 & www.intelligencetest.com/index.htm \\
128 & www.ikea.com/gb/en.html \\
129 & www.nubbeo.com.ar/index.html \\
130 & www.mulberry.com/es/shop/sale/sale-mens-accessories.html \\
131 & www.tous.com/es-es/novedades/relojes/c/59.html \\
132 & preferenceweb.com/collections/all-sneakers.html \\
133 & www.trekbikes.com/us/en\_US/bikes/mountain-bikes/electric-mountain-bikes/c/B512/index.html \\
134 & addons.prestashop.com/es/2-modulos.html \\
135 & us.pandora.net/en/charm-bracelets/pandora-moments/pandora-moments-bracelets/index.html \\
136 & kawaiipenshop.com/index.html \\
137 & www.vam.ac.uk/shop/lindsay-philip-butterfield-blue-flower-silk-scarf.html \\
138 & shop.fendt.com/kids-toys/clothing/shirts.html \\
139 & www.euroholds.com/it/29-prese-arrampicata.html \\
140 & www.emmaclothes.com/index.html \\
141 & www.arduino.cc/en/Main/Software.html \\
142 & naranjascarcaixent.com/tienda.html \\
143 & www.technicalbookstoreonline.com/new-arrivals.php.html \\
144 & www.floridarealestatecollege.com/index.html \\
145 & www.basf.com/nl/nl/who-we-are/BASF-in-Nederland.html \\
146 & www.mcphersonoil.com/index.html \\
147 & www.thirteenhou.com/menu.php.html \\
148 & www.embalajesterra.com/precintadoras-manuales-168.html \\
149 & www.crypto.ch/en/about.html \\
150 & www.shopbookshop.com/index.htm
\\
\hline
\end{tabular}
}
\end{center}
%\label{tbl-paths}
%\caption{Path to the key page of each benchmark}
\end{table}

\subsubsection{Rules for using the suite and report\\}
%\textbf{TODO: Extender un poco con explicacion general y pequeña justificacion de porque estas reglas y no otras}
All researchers and developers that use TECO must follow two basic principles: 
\begin{enumerate}
\item They must publish their results so that they are publicly available.
\item They must provide enough information so that anyone can easily duplicate their experiments.
\end{enumerate}

%%%%%%%%%%%%%%%%%%%%%%%%%%%%%%%%%%%%%%%%%%%%%%%

\section{Conclusions}\label{sec-Other}

This paper presents a benchmark suite composed of 150 heterogeneous websites.
This benchmark suite can be used to test any technique that works with webpages, but it is specially useful for template detection, content extraction and menu detection because it includes a gold standard for them. 
Concretely, the gold standard identifies for each benchmark the template, the main content and the main menu. Thus, it can be used to evaluate and compare techniques and implementations of these disciplines. 
The suite is publicly available and free.

%%%%%%%%%%%%%%%%%%%%%%%%%%%%%%%%%%%%%%%%%%%%%%%

%\begin{thebibliography}{}
%
%
%\end{thebibliography}
\bibliography{\biblio{biblio}}
\bibliographystyle{plain}

\end{document}